\documentclass[12pt]{article} 

\newenvironment{pf}{\noindent{\bf Proof:}}{\newline\kbn}

\usepackage{amssymb,theorem}
\usepackage{german,a4,isolatin1}
\newtheorem{satz}{Theorem}
\newtheorem{folge}[satz]{Corollary}

\newtheorem{prop}[satz]{Proposition}
\newtheorem{defi}[satz]{Definition}

\newcommand{\Einsop}{\leavevmode{\rm 1\mkern  -4.4mu l}}

\newcommand{\lie}[1]{{\mathfrak #1}}
\newcommand{\komm}[2]{{\left[ #1 , #2 \right]}} 
\newcommand{\norm}[1]{{\left\| #1 \right\|}} 
\newcommand{\klammer}[1]{{\left( #1 \right)}}
\newcommand{\skalar}[2]{{\left\langle #1 , #2 \right\rangle }}
\newcommand{\lok}[1]{{\mathcal #1}}

\newcommand{\Name}[1]{{\sc #1}} 
\newcommand{\dopp}[1]{{\mathbb #1}}

\newcommand{\kbn}{$\square$}

\begin{document}

\title{Conformal Covariance Subalgebras}
\author{S\o{}ren K"oster\\Inst.\ f.\ Theor.\ Physik\\ Universit\"at
  G\"ottingen\\ Tammannstr. 1, 37077 G\"ottingen \\ Germany}
\date{} 

\maketitle
\begin{abstract} 
We give a direct \Name{Lie} algebraic characterisation of conformal
inclusions of chiral current algebras associated with compact,
reductive \Name{Lie} algebras. We use quantum field
theoretic arguments and prove a longstanding conjecture of
Schellekens and Warner on grounds of unitarity and positivity
of energy. We explore the structures found to characterise
{\em conformal covariance subalgebras} and {\em coset current
algebras}.\\[1mm]
\end{abstract}

\section{Introduction}
\label{sec:intro}

Conformal inclusions of chiral current algebras are of interest for a
large variety of reasons. Their classification was undertaken some time
ago, because they are particularly relevant to string theory for their
making string compactification possible without altering 
conformal covariance. Using general
arguments this task was transferred to checking maximal inclusions of reductive
\Name{Lie} algebras in simple \Name{Lie} algebras, for which a
classification was available already. The classification of conformal
inclusions was thus achieved, looking at the central charge of the
respective stress-energy tensors, by several authors \cite{AGO87,
  BB87, SW86}.

Many of the conformal inclusions were found to correspond to
{\em symmetric spaces} (cf.\ \cite{GNO85, cD96} in particular), and
{\em isotropy irreducibility} of the coset space proved a useful yet
neither necessary nor 
sufficient criterion for an inclusion being conformal. We undertake a complete
characterisation of conformal inclusions by means of straightforward
arguments familiar in (axiomatic) quantum field theory. On the course
we prove a longstanding\footnote{To the author's surprise, there
  does not seem to be a proof available yet.}
conjecture of \Name{Schellekens and Warner} 
\cite{SW86}. 

We use properties of any  \Name{Wightman} quantum field
theory\footnote{\Name{Wightman}'s axioms are manifestly fulfilled for
  all current algebras which are available as quark models \cite{BH71}
(corresponding to abelian \Name{Lie} algebras $\dopp{R}^n$ and to the
classical \Name{Lie} algebras  of type $A_n$, 
$B_n$, $C_n$, $D_n$ and to the exceptional \Name{Lie} algebra
$G_2$). For the remaining four cases (corresponding to the exceptional
\Name{Lie} algebras  of type $E_6$, $E_7$,
 $E_8$, $F_4$) \Name{Wightman}'s axioms appear implicitly in the
 literature \cite{vKbook}, \cite{GW84}, \cite{vL97}.}:
positivity of energy, separating property of 
the vacuum for local quantum fields, and unitarity. 
Our analysis clarifies the situation in natural
group theoretical terms and in direct correspondence to quantum field
theoretical notions. Moreover, there is no need to specialise in
maximal subalgebras and our approach is rather direct in that respect.

The point of view taken in this work arose  from a more general question:
how does the inner-implementing representation $U^\lok{A}$, uniquely
associated with every 
covariant subtheory $\lok{A}$ of a chiral conformal theory
$\lok{B}$ by means of the \Name{Borchers-Sugawara} construction
\cite{sK02}, act on the observables of the larger theory 
$\lok{B}$?  An answer to the more general question is given
in an independent work \cite{sK03c}.      

We proceed as follows: In the next section we introduce notations
and conventions, prove the conjecture of \Name{Schellekens and Warner}
and provide a direct argument for conformal inclusions being
necessarily restricted to level $1$. The third section is about
studying {\em conformal covariance subalgebras} associated to
\Name{Lie} algebra inclusions, these being intermediate to the
original inclusion, if not trivial. The section will be closed by a
simple characterisation of {\em coset currents}, i.e. current
subalgebras commuting with  the given current subalgebra.


\section{Characterisation of conformal inclusions}
\label{sec:char}

We study a current algebra associated with a simple\footnote{General reasoning leads to an extension of the following
discussion to 
inclusions of reductive subalgebras in reductive \Name{Lie}
algebras, cf.\ eg.\ \cite{AGO87}.}, compact \Name{Lie}
algebra $\lie{g}$ consisting of symmetric operator
valued distributions, the currents. We will treat these as fields on
the chiral light ray. Basis elements of $\lie{g}$ will
be denoted by $T^a$; they give the {\em colour} of the corresponding
current $j^a$. The current algebra is given by the following
commutation relations: 
\begin{displaymath}
  \komm{j^a(x)}{j^b(y)} = i f^{ab}{}_c j^c(x) \delta(x-y) + k
  g^{ab}_\lie{g} \frac{i}{2\pi} \delta' (x-y) \,\, . 
\end{displaymath}
$g_\lie{g}$ denotes the \Name{Killing} metric of $\lie{g}$,
$f^{ab}{}_c$ its structure constants and $k$ the current algebra's
{\em level}; $k$ is a positive integer.

By embedding a reductive \Name{Lie} subalgebra $\lie{h}$ into
$\lie{g}$ via an injective homomorphism $\iota: \lie{h}
\hookrightarrow \lie{g}$ we have an associated current
subalgebra. $\lie{h}$ consists of 
several simple ideals, denoted (for the time being) by
$\lie{h}_\alpha$, and an abelian 
ideal of dimension $n\geq 0$. The inclusions $\iota(\lie{h}_\alpha)
\subset \lie{g}$ are partly characterised by their
\Name{Dynkin} index $I_\alpha$. Denoting the \Name{Killing} metric of
$\lie{h}_\alpha$ by $g_\alpha$ we have the following commutation
relations for currents associated with colours in $\iota(\lie{h}_\alpha)$:
\begin{displaymath}
  \komm{j^{\iota(a)}(x)}{j^{\iota(b)}(y)} = i f^{\iota(a)\iota(b)}{}_{\iota(c)}
    j^{\iota(c)}(x) \delta (x-y) + I_\alpha k g^{ab}_\alpha
    \frac{i}{2\pi} \delta' (x-y) \,\, . 
\end{displaymath}

The infinitesimal conformal transformations are implemented by the
adjoint action of the \Name{Sugawara} stress-energy tensor
$\Theta^\lie{g}$. It is given by:
\begin{displaymath}
   \Theta^\lie{g} (x) = \frac{\pi}{k+v^\lie{g}}
  g^\lie{g}_{ab} :j^{a} j^{b}:(x) \,\, . 
\end{displaymath}
$v^\lie{g}$ is the dual \Name{Coxeter} number of $\lie{g}$. The
commutation relation of $\Theta^\lie{g}$ with a current reads as:
\begin{displaymath}
  \komm{\Theta^\lie{g}(x)}{j^c(y)} = i j^c(x) \delta'(x-y) \,\, . 
\end{displaymath}

Restricting to colours in $\iota(\lie{h}_\alpha)$ the same
construction yields a stress-energy
tensor $\Theta^\alpha$ having the same commutation relation with
currents associated with colours in $\iota(\lie{h}_\alpha)$:
\begin{displaymath}
    \Theta^\alpha (x) =\frac{\pi}{I_\alpha k+v^\alpha}
  g^\alpha_{ab} :j^{\iota(a)} j^{\iota(b)}:(x) \,\, . 
\end{displaymath}

For the abelian ideal we adopt the following conventions:
$I_{\dopp{R}^n}:=1$, $v^{\dopp{R}^n} := 0$, $g_{\dopp{R}^n}^{ij} :=
g^{\iota(i)\iota(j)}_{\lie{g}}$. Using these as input all the formulas
above apply to currents associated with colours in
$\iota(\dopp{R}^n)$. We drop the distinction between simple and
abelian ideals of $\lie{h}$ 
and use the symbol $\lie{h}_\alpha$ for any simple or the abelian
ideal from now on.

With this general notation the action of a stress-energy tensor
$\Theta^\alpha$ on an arbitrary current $j^c$ reads as:
\begin{eqnarray}
\komm{\Theta^\alpha(x)}{j^c(y)}\nonumber
&=&\frac{\pi}{I_\alpha k+ v^\alpha}
 \, g^\alpha_{ab} \,\, i f^{\iota(b)c}{}_d \,\, 
:j^{\iota(a)}j^d+j^dj^{\iota(a)}:(x) \,\, \delta(x-y)\nonumber\\
&&+i\frac{k}{I_\alpha k+ v^\alpha}   \,\,  j^{\iota(a)}(x)  \,\, g^\alpha_{ab}
g_\lie{g}^{\iota(b)c}  \,\, \delta'(x-y)
\nonumber\\&&
+i \frac{1}{2(I_\alpha k+ v^\alpha)}   \,\, j^d(x) 
\klammer{C_2^{\alpha}}_d{}^c  \,\, \delta'(x-y) \label{geneitcomm} \,\, . 
\end{eqnarray}
 This equation is obtained by applying the current algebra
 and the normal ordering prescription for currents \cite{FST89}
. The
 matrix $C_2^{\alpha}$ stands for the second \Name{Casimir} element of
 $\lie{h}_\alpha$ in the representation $Ad_\lie{g}\circ
 \iota|_{\lie{h}_\alpha}$, if $\lie{h}_\alpha$ is a simple ideal. In
 any case we have:
\begin{displaymath}
  \klammer{C_2^{\alpha}}_d{}^c = g^\alpha_{ab} \,\, 
  if^{\iota(b)e}{}_d   \,\, if^{\iota(a)c}{}_e = g^\alpha_{ab}\klammer{
    Ad_{T^{\iota(b)}} Ad_{T^{\iota(a)}} }_d{}^c \,\, . 
\end{displaymath}
Taking the trace of this matrix one may readily see that it does
not vanish for the abelian ideal.

Now we are prepared to state and prove our main
result. \Name{Schellekens and Warner} conjectured it in their
discussion closing \cite{SW86}.
\begin{satz}\label{schwacon}
  The following holds true for the {\bf weighted {Casimir} element}
  $\widetilde{C}_2^{\iota(\lie{h})}$ 
  of $\iota(\lie{h})$ ($P_\alpha$ stands for the projection onto
  $\iota(\lie{h}_\alpha)$): 
  \begin{equation}\label{schwainq}
    \widetilde{C}_2^{\iota(\lie{h})} := \sum_\alpha \frac{2 I_\alpha k P_\alpha + C_2^{\alpha}}{2(I_\alpha
      k+ v^\alpha)} \leqslant \Einsop \,\, . 
  \end{equation}

This inequality is saturated if and only if
$\iota(\lie{h})\subset\lie{g}$ yields a 
conformal inclusion, i.e. $\sum_\alpha \Theta^\alpha =:
\Theta^\lie{h} = \Theta^\lie{g}$. 
\end{satz}

\begin{pf}
By invariance of $g^\lie{g}$ the orthocomplementation $\lie{g}=
\iota(\lie{h}) + \iota(\lie{h})^\perp$ provides a reduction of the
representation $Ad_\lie{g}\circ\iota$. We
have  $C^\alpha_2|_{\iota(\lie{h})} = 2 v^\alpha P_\alpha$,
i.e. $\widetilde{C}_2^{\iota(\lie{h})}|_{\iota(\lie{h})}=\Einsop$, and
the inequality only remains to be proven for colours orthogonal to
$\iota(\lie{h})$, where $P_\alpha|_{\iota(\lie{h})^\perp}=0$. Because
all \Name{Casimir} elements commute and all are 
positive operators, we assume as well that $T^c$ is a common
eigenvector for all linear mappings $C^\alpha_2$.

We prove the inequality by looking at specific expectation values of the
coset \Name{Hamilton}ian $L_0^\lie{g}-L_0^\lie{h}$. This is a positive
operator, which is given 
by the coset stress-energy tensor
$\Theta^\lie{g}-\Theta^\lie{h}$ smeared with the test function
$\xi_{L_0}(x)= \frac{1}{2}(x^2+1)$. The infinitesimal action of a
conformal \Name{Hamilton} operator on the test function of a smeared
field covariant with respect 
to it shall be abbreviated by $l_0$, i.e. we have
$$\komm{L_0^\lie{g}}{j^c(g)} = i \int dx g'(x) \xi_{L_0}(x) j^c(x)\equiv i \int
dx  \,\, l_0 g (x) \, j^c(x) = i\, j^c(l_0 g) \,\, . $$

Using the general commutation relation (\ref{geneitcomm}), calculating
two and three point functions of currents (cf.\ \cite{FST89}), observing
that some group-theoretical tensors involved are null for reasons of
permutation symmetry/ antisymmetry and carefully taking into
account the normal ordering of currents    \cite{FST89} one arrives at
the following formula:
\begin{eqnarray}
  0 &\leqslant& \langle\Omega , j^c(g)^\dagger (L_0^\lie{g}-L_0^\lie{h})
    j^c(g)\Omega\rangle\nonumber\\
&=&i \klammer{1- \sum_\alpha \frac{C^\alpha_2[T^c]}{2(I_\alpha k+
    v^\alpha)}} \langle\Omega , j^c(g)^\dagger
    j^c(l_0 g)\Omega\rangle\label{star} \,\, . 
\end{eqnarray}
The desired inequality may be established through
  division by $i\langle\Omega,j^c(g)^\dagger
    j^c(l_0 g)\Omega\rangle$, which does not vanish for generic $g$ and is
  positive as an expectation value of $L_0^\lie{g}-L_0^\lie{h}\geq 0$. 

If we have
$\Theta^\lie{g}=\Theta^\lie{h}$,  (\ref{schwainq}) is
saturated on $\iota(\lie{h})$ trivially and because of (\ref{star}) on
$\iota(\lie{h})^\perp$ as well, hence on all of $\lie{g}$. The
conclusion in the opposite direction  
is, actually, a consequence of equation (\ref{eq:geneittwo}) in
proposition \ref{crueqns}
: This leads to trivial commutation relations for
$\Theta^\lie{g}-\Theta^\lie{h}$, especially to $c_\lie{g}=c_\lie{h}$,
which yields, by a variant of the
 \Name{Reeh-Schlieder} theorem\footnote{\label{foot:RS} See for
   example \cite{rJ65} (lemma 2, 
 section V.3.B); for an argument directly referring to the
 \Name{Virasoro} algebra see \cite{GW85},\cite{fG86a}.},
 $\Theta^\lie{g}-\Theta^\lie{h} = 0$. 
\end{pf} 


\begin{folge}\label{kone}
  An embedding $\iota(\lie{h})\subset \lie{g}$ can give rise to a
  conformal inclusion of the associated current algebras only, if the
  current algebra associated with $\lie{g}$ has level $k=1$.
\end{folge}

\begin{pf}
Highest-weight representations of current algebras may be characterised
uniquely by a vector of lowest energy which is a highest-weight vector with respect to the {\em horizontal subalgebra} of
currents $j^a([1])$ smeared with the constant testfunction $[1](x)=1$.
We look at the representation defined by the highest weight
$\psi_\lie{g}$ of the adjoint representation of $\lie{g}$. Since
$\psi_\lie{g}$ has, by the usual convention, length 2, this
representation is in accordance with the \Name{Weyl}-alcove condition
\cite[(4.51)]{FST89} for unitary representations of current algebras
for $k\geq 2$. The following argument applies, therefore, to  all but
level $1$. 

Actually, we may restrict attention to the action of
$L_0^\lie{g}-L_0^\lie{h}$ on $\lie{g} \psi_\lie{g}$, the
highest-weight module of $\lie{g}$ generated from the vector with
lowest energy and highest weight $\psi_\lie{g}$. Here we have:
\begin{displaymath}
  0 \leqslant \left.\klammer{L_0^\lie{g}-L_0^\lie{h}}\right|_{\lie{g}
    \psi_\lie{g}} = \frac{v^\lie{g}}{k+v^\lie{g}} \Einsop - \sum_\alpha
  \frac{C^\alpha_2}{2(I_\alpha k+ v^\alpha)} \,\, . 
\end{displaymath}
This implies a strictly sharper bound than inequality (\ref{schwainq})
and by theorem 
\ref{schwacon} immediately yields the desired result.
\end{pf}


\section{Covariant and invariant colours}
\label{sec:cocoalg}

After we have given a characterisation of conformal inclusions
$\iota(\lie{h})\subset\lie{g}$, we now pursue further the structures in
colour space which are associated with the action of $\Theta^\lie{h}$
on currents with colours in $\lie{g}$. We find that {\em covariant} and
{\em invariant} colours form reductive \Name{Lie} algebras,
the first being intermediate to the original embedding
$\iota(\lie{h})\subset\lie{g}$, the second being orthogonal to and
commuting with it. All these results are in terms of the {\em weighted
{Casimir} element} $\widetilde{C}_2^{\iota(\lie{h})}$ of the
\Name{Lie} algebra $\iota(\lie{h})$, 
which already appeared in the previous section.

The following is the main ingredient of the results in this section:
\begin{prop} \label{crueqns}
For an arbitrary colour $T^c\in\lie{g}$ we have:
  \begin{eqnarray}
    \label{eq:geneittwo}
&&\norm{\komm{\klammer{\Theta^\lie{g}-\Theta^\lie{h}}(f)}{j^c(g)}\Omega}^2
\nonumber\\ 
&=&8k \pi^2 \langle(\Einsop - \widetilde{C}_2^\lie{h})
  T^c , \widetilde{C}_2^\lie{h} T^c\rangle{}_\lie{g}
\,\,\widetilde{\Delta}^4(\overline{f\cdot g}, f\cdot g) \nonumber\\
&& + 
k
\langle(\Einsop-\widetilde{C}^\lie{h}_2)T^c ,
(\Einsop-\widetilde{C}^\lie{h}_2)T^c\rangle_\lie{g}  
\,\,\widetilde{\Delta}^2 (\overline{f\cdot g'}, f\cdot g')  \,\, . 
  \end{eqnarray}
Here $\skalar{.}{.}_\lie{g}$ stands for the scalar product on
$\lie{g}$ induced by the \Name{Killing} form. We define 
$$\Phi^c(x) := \sum_\alpha  
\frac{1}{2(I_\alpha k + v^\alpha)} g^\alpha_{ab} f^{\iota(b)c}{}_d
:j^{\iota(a)}j^d+j^dj^{\iota(a)}:(x) \,\, . $$
The two-point function of $\Phi^c$ is given by:
\begin{equation}
  \label{eq:covfietwo}
  \langle\Omega , \Phi^c(x)\Phi^c(y)\Omega\rangle = 2 k
  \widetilde{\Delta}^4 (x-y) \langle(\Einsop - \widetilde{C}_2^\lie{h})
  T^c , \widetilde{C}_2^\lie{h} T^c\rangle{}_\lie{g} \,\, . 
\end{equation}

The numerical distributions in these formulae are given by:
\begin{eqnarray*}
  \widetilde{\Delta}^4(\overline{f\cdot g}, f\cdot g) &=&(2\pi)^{-4}
  \int\!\!\!\int dx \,dy \,\,\klammer{i[(x-y)-i\varepsilon]}^{-4} \overline{f\cdot g}(x)f\cdot g(y) \,\, , \\
 \widetilde{\Delta}^2(\overline{f\cdot g'}, f\cdot g') &=&(2\pi)^{-2}
  \int\!\!\!\int dx \, dy\,\, \klammer{i[(x-y)-i\varepsilon]}^{-2} \overline{f\cdot g'}(x)f\cdot g'(y) \,\, . 
\end{eqnarray*}
\end{prop}

\begin{pf}
  We will not give the derivation of these formulae in detail. We
  rather indicate their verification. First, one may restrict
  attention to colours $T^c\in\iota(\lie{h})^\perp$ since the weighted
  \Name{Casimir} respects the orthogonal decomposition
  $\lie{g}=\iota(\lie{h}) + \iota(\lie{h})^\perp$ with respect to
  $Ad\circ\iota$ and $\Theta^\lie{g}-\Theta^\lie{h}$ commutes with all
  currents whose colours are in $\iota(\lie{h})$. ``All'' that one has
  to do  is to   
  apply the general commutation relation (\ref{geneitcomm}) restricted
  to colours from $\iota(\lie{h})^\perp$, follow carefully the normal
  ordering of currents, observe symmetries of group theoretical
  coefficients, keep in mind $T^c\in\iota(\lie{h})^\perp$, calculate
  some $n$-point functions of currents 
  following the scheme in \cite{FST89}, use \Name{Jacobi}'s identity a
  few times and recognise the second \Name{Casimir} element in the
  adjoint representation, which amounts to twice the dual
  \Name{Coxeter} number. With all that, it is a straightforward
  algebraic exercise. 
\end{pf}

{\em Remark:} Taking  $g$ as the test function of
constant value $1$, equation (\ref{eq:geneittwo}) implies
$\widetilde{C}_2^\lie{h} (\Einsop -\widetilde{C}_2^\lie{h})\geq 0$,
from which we immediately get inequality (\ref{schwainq}), and the second
statement in  theorem \ref{schwacon} follows from
(\ref{eq:geneittwo}), too.

\begin{defi}
  A current $j^c$ is said to transform covariantly  with respect to
  $\Theta^\lie{h}$, if and only if $\komm{\Theta^\lie{g}(f)}{j^c(g)} =
  \komm{\Theta^\lie{h}(f)}{j^c(g)}$ $\forall f,g$.
\end{defi}

\begin{folge}\label{confcoval}
  A current $j^c$ transforms covariantly  with respect to
  $\Theta^\lie{h}$, if and only if its
  colour fulfills: 
  $\widetilde{C}_2^{\iota(\lie{h})} T^c = T^c$. These {\bf covariant
  colours} form a reductive \Name{Lie}
  algebra, $\lie{k}$, containing $\iota(\lie{h})$ as a subalgebra. If
  $\lie{k}\neq \iota(\lie{h})$, then the level of the current algebra
  associated with $\lie{g}$ has to be $k=1$.  
\end{folge}

\begin{pf}
If we have $\widetilde{C}_2^{\iota(\lie{h})} T^c = T^c$, we know from
the variant of the \Name{Reeh-Schlieder} theorem (see footnote \ref{foot:RS})
and proposition \ref{crueqns} above, that $j^c$ and 
the coset stress-energy tensor commute. This is another way of saying:
$j^c$ transforms covariantly with respect to 
  $\Theta^\lie{h}$. 

Conversely: If $j^c$ is covariant with respect to
$\Theta^\lie{h}$, the group theoretical scalar products  in
equation (\ref{eq:geneittwo}) have to be zero, since the numerical
distributions involved are linearly independent. The second one of
these is the norm of $(\Einsop -\widetilde{C}_2^{\iota(\lie{h})})T^c$,
which makes the equation $\widetilde{C}_2^{\iota(\lie{h})} T^c =T^c$
valid.
Now, if $T^a$ and $T^b$ are covariant colours, then so is $-i
[T^a, T^b]$. This is clear, if one observes
$f^{ab}{}_c j^c(g) = -i [j^a([1]), j^b(g)]$, where $[1]$ is a
constant test function: $[1](x) = 1$. 

The reductivity of $\lie{k}$ is not difficult to prove, either. 
$\lie{k}$
is a subspace
of $\lie{g}$, endowed with an invariant scalar product, which is given by the
restriction of the \Name{Killing} form on $\lie{g}$. By invariance of
this scalar product on $\lie{k}$ with respect to $Ad_\lie{k}$ (this
being a mere restriction of invariance under $Ad_\lie{g}$) any
invariant subspace of $\lie{k}$ has an invariant orthogonal
complement. Now this is complete reducibility of $Ad_\lie{k}$ and thus
 $\lie{k}$ is reductive \cite{jC89} (25.3.a).

Since one can reduce the problem of understanding all conformal
inclusions to the studies of reductive inclusions in simple \Name{Lie}
algebras (cf.\ eg.\ \cite{AGO87}) the last part follows immediately from
corollary  
\ref{kone}, as $\iota(\lie{h})\subset \lie{k}$ is, by construction of
$\lie{k}$, a conformal inclusion and \Name{Dynkin} indices of the
simple ideals in $\lie{k}$ are greater than or equal to $1$.
\end{pf}

\begin{folge}
  A current $j^c$, whose colour $T^c$ lies in $\iota(\lie{h})^\perp$
  and fulfills $\widetilde{C}_2^{\iota(\lie{h})} T^c = 0$, commutes
  with the entire current algebra associated with
  $\iota(\lie{h})$. These colours form a reductive \Name{Lie} algebra,
  the algebra of {\bf invariant colours}; we call their
  current algebra {\bf coset current algebra}.  
\end{folge}

\begin{pf}
 We set
  $V_0:=  ker(\widetilde{C}_2^{\iota(\lie{h})}) \cap
  \iota(\lie{h})^\perp$. $V_0$  is an invariant  subspace with respect
  to  the action of $\lie{h}$ on $\lie{g}$ via  
$Ad_\lie{g}\circ\iota$. In fact, it is the representation 
space for the trivial subrepresentation on $\iota(\lie{h})^\perp$: For
any simple ideal 
$\lie{h}_\alpha$ we have by complete reducibility $C_2^\alpha|_{V_0} =
\sum_\Lambda \langle\Lambda+ 2 \rho , \Lambda\rangle_{\lie{h}_\alpha} =
0$. Since both the \Name{Weyl} vector $\rho$ and the contributing
highest-weight vectors $\Lambda$ are dominant, we have $\Lambda =
0$. For the 
abelian ideal the irreducible subrepresentations on $V_0$ are given by
common eigenvectors, such that $C_2^{\dopp{R}^n} v =
g^\lie{g}_{\iota(i)\iota(j)} \lambda^i\lambda^j v = 0$. This gives the
same result. This means, that all of $V_0$ commutes with
$\iota(\lie{h})$, i.e. $V_0\subset \iota(\lie{h})'$. We gain directly:
$V_0 = \iota(\lie{h})' \cap \iota(\lie{h})^\perp$.

By \Name{Jacobi}'s identity and invariance of the \Name{Killing}
metric, $\iota(\lie{h})' \cap\iota(\lie{h})^\perp$ forms a \Name{Lie}
subalgebra of $\lie{g}$. This is
reductive by the same argument as in the proof of corollary
\ref{confcoval}. 
\end{pf}

\noindent{\em Two concluding remarks:} Generically, the coset theory is not
generated by coset currents
. Obviously $\iota(\lie{h})\oplus(\iota(\lie{h})'
\cap \iota(\lie{h})^\perp)\subset \lie{g}$ has to be a conformal
inclusion for that to be the case, since the coset stress-energy
tensor has to be the \Name{Sugawara} stress-energy tensor of the
current algebra associated with $\iota(\lie{h})'
\cap \iota(\lie{h})^\perp$. \Name{Casimir} elements of $\iota(\lie{h})'
\cap \iota(\lie{h})^\perp$ give, when transferred to the corresponding
horizontal subalgebra,  charge operators of the coset
theory. These will, in general, fail to separate the representations
of the coset theory. The same goes for the \Name{Cartan} subalgebra of
$\iota(\lie{h})'\cap \iota(\lie{h})^\perp$, whose spectrum defines
characters of the representations of the coset theory.  The
coset current algebra is trivial for all inclusions with {\em minimal} coset
theory: Here the coset theory is generated by the coset stress-energy
tensor and this theory contains nothing but this field
\cite{sC98}.  Triviality of coset current algebra ought to be regarded
as the generic situation. 

Currents $j^c$ with vanishing covariance field $\Phi^c$ are linear
combinations of covariant and coset currents. This is obvious, since
a decomposition of $T^c$ into eigenvectors
of $\widetilde{C}_2^\lie{h}$ with distinct eigenvalues $\lambda$
yields (cf.\ equation \ref{eq:covfietwo}): 
$$\langle(\Einsop - \widetilde{C}_2^\lie{h}) 
  T^c , \widetilde{C}_2^\lie{h} T^c\rangle{}_\lie{g} = \sum_\lambda \lambda
(1-\lambda) \langle
  T^c_\lambda , T^c_\lambda\rangle{}_\lie{g} \,\, . $$
 As $0\leq \lambda \leq 1$ this scalar 
product vanishes, if and only if just $0$ and $1$
contribute. This means, that there are no currents with a ``simple''
intermediate transformation behaviour with respect to the action of
$\Theta^\lie{h}$. Typically, a current $j^c$ has $\Phi^c\neq 0$, i.e.
a ``complicated'' transformation behaviour. By the analysis in
\cite{sK03c} this behaviour is known to be physically satisfactory, still.  

\subsection*{Acknowledgements}
I thank \Name{K.-H.Rehren} (G\"ottingen) for
  many helpful discussions and a critical reading of the
  manuscript. Financial support from the \Name{Ev.\ Studienwerk 
    Villigst} is gratefully acknowledged.


\begin{thebibliography}{GNO85}
\providecommand{\selectlanguage}[1]{\relax}

\bibitem[AGO87]{AGO87}
R.C. Arcuri, J.F. Gomes, and D.I. Olive.
\newblock {\em Conformal subalgebras and symmetric spaces\/}.
\newblock Nuclear Phys. {\bf B285} (1987) 327.

\bibitem[BB87]{BB87}
F.A. Bais and P.G. Bouwknegt.
\newblock {\em A classification of subgroup truncations of the bosonic
  strings\/}.
\newblock Nuclear Phys. {\bf {} B279} (1987) 561.

\bibitem[BH71]{BH71}
K.~Bardakci and M.~B. Halpern.
\newblock {\em New dual quark models\/}.
\newblock Phys. Rev. {\bf D3} (1971) 2493.

\bibitem[Car98]{sC98}
S.~Carpi.
\newblock {\em Absence of subsystems for the {H}aag-{K}astler net generated by
  the energy-momentum tensor in two-dimensional conformal field theory\/}.
\newblock Lett. Math. Phys. {\bf 45} (1998) 259--267.

\bibitem[Cor89]{jC89}
J.F. Cornwell.
\newblock {\em Group Theory in Physics, Volume {III}: Supersymmetries and
  Infinite-Dimensional Algebras\/}, volume~10 of {\em Techniques in Physics\/}.
\newblock Academic Press, London, San Diego, 1989.

\bibitem[Dab96]{cD96}
C.~Daboul.
\newblock {\em Algebraic proof of the symmetric space theorem\/}.
\newblock J. Math. Phys. {\bf 37} (1996) 3576--3586.

\bibitem[FST89]{FST89}
P.~Furlan, G.M. Sotkov, and I.T. Todorov.
\newblock {\em Two-dimensional conformal quantum field theory\/}.
\newblock Riv. Nuovo Cim. {\bf 12} (1989) 1--203.

\bibitem[GNO85]{GNO85}
P.~Goddard, W.~Nahm, and D.~Olive.
\newblock {\em Symmetric spaces, {S}ugawara's energy momentum tensor in two
  dimensions and free fermions\/}.
\newblock Phys. Lett. {\bf B160} (1985) 111--116.

\bibitem[Gom86]{fG86a}
F.~Gomes.
\newblock {\em The triviality of representations of the {V}irasoro algebra with
  vanishing central element and {$L_0$} positive\/}.
\newblock Phys. Lett. {\bf 171B} (1986) 75.

\bibitem[GW84]{GW84}
R.~Goodman and N.R. Wallach.
\newblock {\em Structure and unitary cocycle representations of loop groups and
  the group of diffeomorphisms of the circle\/}.
\newblock J. Reine Angew. Math. {\bf 347} (1984) 69--222.

\bibitem[GW85]{GW85}
R.~Goodman and N.R. Wallach.
\newblock {\em Projective unitary positive-energy representations of ${D}${\it
  iff}$({S}^1)$\/}.
\newblock J. Funct. Anal. {\bf 63} (1985) 299--321.

\bibitem[Jos65]{rJ65}
R.~Jost.
\newblock {\em The General Theory of Quantized Fields\/}, volume~IV of {\em
  Lectures in Applied Mathematics\/}.
\newblock American Mathematical Society, Providence, RI, 1965.
\newblock Proceedings of the Summer Seminar, Boulder, Colorado, 1960, Marc Kac
  (ed.).

\bibitem[Kac90]{vKbook}
V.~Kac.
\newblock {\em Infinite dimensional {L}ie algebras\/}.
\newblock Cambridge University Press, 3rd edition, 1990.

\bibitem[K{\"{o}}s02]{sK02}
S.~K{\"{o}}ster.
\newblock {\em Conformal transformations as observables\/}.
\newblock Lett. Math. Phys. {\bf 61} (2002) 187--198.

\bibitem[K{\"{o}}s03]{sK03c}
S.~K{\"{o}}ster.
\newblock {\em Local nature of coset models\/}, 2003.
\newblock {m}ath-ph/0303054.

\bibitem[SW86]{SW86}
A.N. Schellekens and N.P. Warner.
\newblock {\em Conformal subalgebras of {K}ac-{M}oody algebras\/}.
\newblock Phys. Rev. {\bf D 34} (1986) 3092--3096.

\bibitem[TL97]{vL97}
V.~Toledano~Laredo.
\newblock {\em Fusion of Positive Energy Representations of
  ${L}{S}pin_{2n}$\/}.
\newblock Ph.D. thesis, University of Cambridge, 1997.

\end{thebibliography}

\end{document}